\documentclass[superscriptaddress,showpacs,amsmath,amssymb,aps,prd,twocolumn,floatfix,nofootinbib, amsart]{revtex4-1}

\usepackage{graphicx}
\usepackage{dcolumn}
\usepackage{amsthm}
\usepackage{bm}
\usepackage{color}
\usepackage{amsmath}
 \usepackage{microtype}
\usepackage{amssymb}
\usepackage{bbm}
\usepackage{hyperref}
\usepackage{todonotes}
\usepackage{hhline}
\hypersetup{
    colorlinks = true
}
\usepackage{lmodern}
\usepackage{textcomp}
\usepackage{etoolbox}
\usepackage{enumitem}
\usepackage{mathrsfs}
\usepackage{bbm}
\usepackage[ruled,vlined]{algorithm2e}
\usepackage{algorithmic}
\usepackage{amsmath}
\usepackage{hyperref}
\usepackage{subfig}
\usepackage{booktabs, tabularx}
\usepackage{siunitx}

\usepackage{array}
\usepackage{cellspace}
\usepackage{natbib}
\newcolumntype{L}[1]{>{\raggedright\arraybackslash}p{#1}}
\newcolumntype{Y}{>{\raggedright\arraybackslash}X} 
\setenumerate[0]{label=$(\alph*)$}
\renewrobustcmd{\bfseries}{\fontseries{b}\selectfont}
\renewrobustcmd{\boldmath}{}
\newrobustcmd{\B}{\bfseries}

\theoremstyle{definition}

\theoremstyle{remark}

\numberwithin{equation}{section}

\newcommand{\norm}[1]{\left\lVert#1\right\rVert}

\newcommand{\bbZ}{\mathbb{Z}}
\newcommand{\bbR}{\mathbb{R}}

\begin{document}

\title{Scalable, Proposal-free Instance Segmentation Network for 3D Pixel Clustering and Particle Trajectory Reconstruction in Liquid Argon Time Projection Chambers}


\newcommand{\SLAC}{SLAC National Accelerator Laboratory, Menlo Park, CA, 94025, USA}
\affiliation{\SLAC}
\newcommand{\ICME}{ICME, Stanford University, Stanford, CA, 94305, USA}
\affiliation{\ICME}
\newcommand{\Stanford}{Stanford University, Stanford, CA, 94305, USA}
\affiliation{\Stanford}

\author{Dae~Heun~Koh} \email{koh0207@stanford.edu} \affiliation{\Stanford}
\author{Pierre~C\^ote~de~Soux} \affiliation{\ICME}
\author{Laura~Domin\'e} \affiliation{\Stanford}
\author{Fran\c cois~Drielsma} \affiliation{\SLAC}
\author{Ran~Itay} \affiliation{\SLAC}
\author{Qing~Lin} \affiliation{\SLAC}
\author{Kazuhiro~Terao} \affiliation{\SLAC}
\author{Ka~Vang~Tsang} \affiliation{\SLAC}
\author{Tracy~L.~Usher} \affiliation{\SLAC}

\collaboration{on behalf of the DeepLearnPhysics Collaboration}\noaffiliation

\begin{abstract}
    Liquid Argon Time Projection Chambers (LArTPCs) are high resolution particle imaging detectors, employed by accelerator-based neutrino oscillation experiments for high precision physics measurements. While images of particle trajectories are intuitive to analyze for physicists, the development of a high quality, automated data reconstruction chain remains challenging. One of the most critical reconstruction steps is particle clustering: the task of grouping 3D image pixels into different particle instances that share the same particle type.  In this paper, we propose  the first scalable deep learning algorithm for particle clustering in LArTPC data using sparse convolutional neural networks (SCNN). Building on previous works on SCNNs and proposal free instance segmentation, we build an end-to-end trainable instance segmentation network that learns an embedding of the image pixels to perform point cloud clustering in a transformed space.
    We benchmark the performance of our algorithm on PILArNet, a public 3D particle imaging dataset, with respect to common clustering evaluation metrics. 3D pixels were successfully clustered into individual particle trajectories with 90~\% of them having an adjusted Rand index score greater than 92~\% with a mean pixel clustering efficiency and purity above 96~\%. This work contributes to the development of an end-to-end optimizable full data reconstruction chain for LArTPCs, in particular pixel-based 3D imaging detectors including the near detector of the Deep Underground Neutrino Experiment. Our algorithm is made available in the open access repository, and we share our Singularity software container, which can be used to reproduce our work on the dataset.
\end{abstract}

\keywords{deep learning;convolutional neural networks;CNNs;instance segmentation;clustering;submanifold convolution;sparse convolution;sparse data;lartpc;scalability}

\maketitle
\section{Introduction}

The present and future accelerator-based neutrino experiments are designed for critical physics measurements including CP-violation~\cite{dune}, search for sterile neutrinos~\cite{sbn}, and rare physics processes such as proton decay and supernova neutrino bursts~\cite{dune,sbn,MicroBooNE}. These measurements require a large-scale, high-precision detector technology and efficient data reconstruction techniques. A Liquid Argon Time Projection Chamber (LArTPC), characterized by its capability in high precision imaging ($\approx$mm/pixel) of charged particles with calorimetric measurement, is the detector technology of choice for many of these experimental programs including the Short Baseline Neutrino (SBN) program~\cite{sbn,MicroBooNE} and Deep Underground Neutrino Experiment~\cite{dune}. The imaging capability of LArTPCs allows one to analyze neutrino interactions in full detail using recorded traces of charged particles traveling inside the liquid argon medium. 

Although LArTPCs produce high resolution images with detailed information on event topology, the remaining challenges include the accurate reconstuction of individual particle information as well as inferring the energy and flavor of neutrinos from 2D and 3D images. Meanwhile, astonishing progress has been made in the computer vision community in the past decade, where deep learning algorithms have been demonstrated to achieve near human level accuracy in image classification, object detection, and semantic segmentation. As a result, deep learning algorithms have found universal applicability in multiple domains that require abstract geometrical reasoning. 
The success of deep learning in typical computer vision tasks naturally led to its application in processing LArTPC data~\cite{UBPaper1,UBPaper2}. In particular, development of a fast and automated reconstruction chain is especially crucial for accelerator-based neutrino experiments that exploit high event rates for accurate determination of oscillation parameters and the measurements of neutrino-nucleus cross-sections. 

\subsection{Particle Clustering in LArTPC Event Data}

\begin{figure*}[t]
    \centering
    \includegraphics[width=\textwidth]{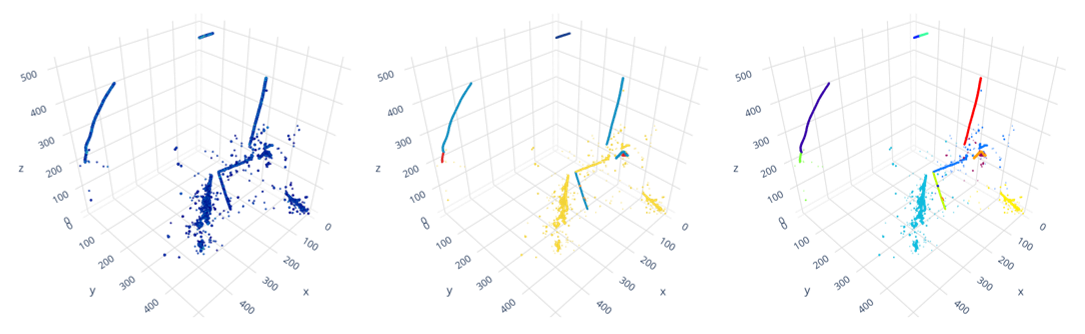}
    \caption{Example event showing energy depositions (left), true semantic labels (center), and true instance labels (right).}
    \label{fig:labels}
\end{figure*}

In this paper, we present a deep convolutional neural network architecture for particle clustering of LArTPC images. \emph{Particle clustering} refers to a task of assigning each and every pixel with a unique instance label so that pixels originating from an ionization trace of a single particle are grouped into the same instance. For instance, accurately sorting each pixel into its corresponding instance is crucial for the identification of interaction vertices or the accurate reconstruction of kinematic variables for each particle. In computer vision, the task of partitioning the image according to each distinct instance is called \emph{instance segmentation}. Therefore, we shall use the terms particle clustering and particle instance segmentation interchangeably, unless otherwise stated. In later sections, we will train our neural network architecture on a Monte Carlo simulation and quantify the performance of each method with respect to common evaluation metrics related to instance segmentation. 

The study presented in this paper is fully reproducible using a \textsc{Singularity}~\cite{Singularity} software container \footnote{\href{https://singularity-hub.org/containers/11757}{https://singularity-hub.org/containers/11757}}, implementations available in the \texttt{lartpc\_mlreco3d}\footnote{\href{https://github.com/DeepLearnPhysics/lartpc_mlreco3d}{https://github.com/DeepLearnPhysics/lartpc\_mlreco3d}} repository and public simulation samples~\cite{PublicSample} made available by the DeepLearnPhysics collaboration.

\subsection{Related Work}

Semantic segmentation may be defined as a pixel-wise classification task. Given a set of pixels of an image, the algorithm is trained to predict a category for each pixel. Instance segmentation, on the other hand, is the task of uniquely labeling each instance within a given semantic class separately. Figure~\ref{fig:labels} illustrates the different tasks in the context of LArTPC events. With the recent progress in deep learning, Convolutional Neural Networks (CNNs)~\cite{lecun_cnn} have become the standard method of solving semantic and instance segmentation tasks, and landmark CNN architectures such as Fully-Convolutional Networks (FCN)~\cite{fcn}, U-Net~\cite{unet}, and Mask-RCNN~\cite{mask_rcnn} have found wide range of applications ranging from biomedical segmentation~\cite{unet} to particle physics experiments~\cite{Radovic2018}. 

Let us define a given instance segmentation architecture to be \emph{proposal-free} if generating instance labels does not require region proposals, which are often realized as bounding box regressions that capture the spatial extent of an object. In the following sections, we shall demonstrate the need for \emph{proposal-free} instance segmentation and why traditional instance segmentation methods with regions proposals may not be ideal for segmenting LArTPC events. Then, building on the recent work by~\cite{laura} and CNN-based instance segmentation methods~\cite{disc,spatial_embeddings}, we implement a \emph{proposal-free} instance segmentation network with 3D \emph{submanifold sparse convolutional neural networks}~\cite{submfdcnn, 3dsubmfdcnn} for LArTPC particle clustering.

\subsubsection{Bounding-box Instance Segmentation Methods}
Bounding-box instance segmentation methods, also called \emph{region-proposal based} instance segmentation methods, employ a two stage process. First, the object detection branch of the network isolates an instance by drawing a bounding box around each instance, and the mask generating branch performs a semantic segmentation task inside each region proposal. Mask-RCNN~\cite{mask_rcnn} is certainly the most popular architecture under the bounding-box instance segmentation category. 

\begin{figure}[t]
    \centering
    \includegraphics[width=0.48\textwidth]{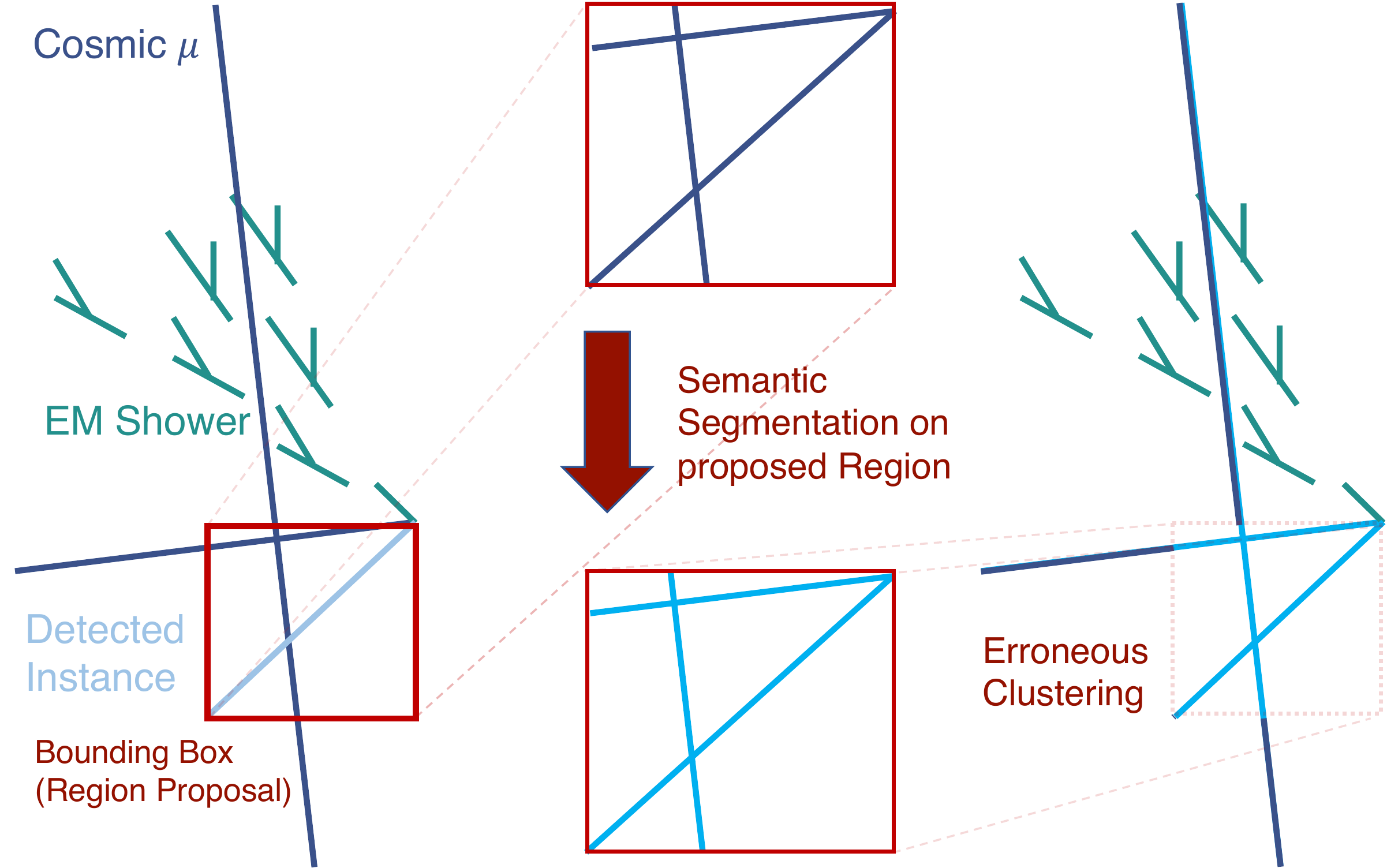}
    \caption{An illustration of an occlusion issue for region-proposal based instance segmentation.}
    \label{fig:occlusion_example}
\end{figure}

A major drawback for particle clustering using region-proposal based architectures is in instance occlusion. Since these methods reduce the instance segmentation task into a semantic segmentation task on a confined volume, if two different instances occupy similar bounding boxes and have similar shape, it becomes challenging to distinguish pixels that belong to the same class but nonetheless belong to separate instances. This is particularly true for simple geometry such as a track-like particle trajectory, essentially a 1D line, compared to more complex objects in photographs such as humans and animals. For instance, consider the diagram of Figure \ref{fig:occlusion_example}. Due to the geometric arrangement of the muon tracks, separating the three instances is a trivial task for a human observer (who even, need not be a domain expert). Yet region proposal based neural network architectures can achieve zero loss while segmentating the whole set of muon pixels as belonging to one instance. As a result, these methods often perform poorly on datasets that contain considerable amount of occlusions between different instances with the same semantics. This is especially problematic for LArTPC events (perhaps more so in 2D images), as in almost every event we expect to see some amount of instance occlusions, such as intersecting tracks and overlapping electromagnetic (EM) shower fragments.
Note also that, in principle, the bounding-box method does not place any restriction on pixels sharing different instance labels. 

This structural limitation of bounding box methods makes their direct application to LArTPC data unappealing. Of course, it may be possible to introduce additional heuristics, post-processing methods, or fine-tuning of both the architecture and the loss to address the aforementioned problems in the detect-and-segment paradigm. Further research on making region proposal based instance segmentations methods to effectively handle such boundary cases may provide useful algorithms directly applicable to LArTPC events. However, we are pursuing an alternative path and require our algorithm to be inherently \emph{proposal-free}. 

\subsubsection{Proposal Free Instance Segmentation Methods}

Let $I \subset \bbZ^3 \times \bbR$ be a 3D image with integer coordinates and one real-valued feature. A common strategy among proposal-free instance segmentation architectures is to learn a coordinate transform function $f: I \to \bbR^d$ to a $d$-dimensional embedding space. For example, the coordinate transform function $f$ may be parameterized by a neural network which is trained under an embedding loss $\mathcal{L}_{emb}$ similar to a k-means objective~\cite{disc}. Here, the pixel-level features represent the coordinates in $\bbR^d$, where points are grouped closer together if they share the same instance label. The assumption behind this  construction~\cite{disc} is that clustering different instances can be done in the embedding space more easily using algorithms such as density-based spatial clustering of applications with noise (DBSCAN) \cite{dbscan} and mean-shift \cite{mean_shift}. Although the idea of learning an embedding map that generates clusters of pixels in $\bbR^d$ is appealing, performance of such methods often falls short of usual region-proposal based methods due to the disconnect between the optimization objective and the label generating process. 


However, there has been considerable progress in search for efficient proposal-free instance segmentation methods as a way to avoid some of the common shortcomings of the region-proposal based methods. One complication among proposal-free methods is their scalability with respect to the number of (nonzero) pixels in a given image. Models such as SGPN~\cite{sgpn} or \emph{recurrent pixel embeddings}~\cite{recurrent_embeddings} construct a similarity matrix of the pixel embeddings and defines a pairwise loss to a group of similar pixels to one another. Yet as constructing the similarity matrix requires $O(N^2)$ memory for $N$ number of pixels, such methods are unlikely to be scalable in the context of processing LArTPC data. Other methods, such as Dynamic Graph CNNs (DGCNN)~\cite{dgcnn} employ k-nearest neighbor (kNN) queries to reason contextual information for point cloud data; yet these methods require kNN queries during both training and inference, which causes a comparatively longer inference time than other instance segmentation networks. 

In this note, we consider sparse convolutional neural network (SCNN) implementations of the method outlined in Ref.~\cite{spatial_embeddings}, which enables the network to generate embeddings that directly maximize the intersection-over-union measure of each instance mask. Since LArTPC data is inherently sparse with less than 1\% of pixels being non-zero, SCNNs are especially well-suited for designing computationally scalable convolutional neural networks for LArTPC applications, as was demonstrated previously~\cite{laura}.  We explain the architecture details in the following sections.

\subsection{Dataset}

Before proceeding to implementation details, we first establish precise definitions of semantic and instance segmentation labels used throughout training and performance evaluation. The open dataset~\cite{PublicSample} used in this study consists of five semantic labels with the following conventions for assigning distinct particle identification labels:
\begin{enumerate}
    \item Heavily Ionizing Particle (HIP)
    
    \item Minimal Ionizing Particle (MIP)
    
    \item EM Shower (Shower)
    
    \item Delta-rays (Deltas)
    
    \item Michel Electrons (Michel)
\end{enumerate}
The instance label for each particles is obtained from the particle id number defined by the GEANT4 simulation~\cite{Geant4}, except for EM showers, where we define a single EM shower particle by grouping all pixels that originate from a common primary ionization. 

\section{Network Architecture}

\subsection{Sparse UResNet}

In designing convolutional neural network architectures for LArTPC event images, we construct models with \emph{submanifold sparse convolutions}~\cite{submfdcnn, 3dsubmfdcnn} instead of dense convolution operations in conventional architectures. The motivation behind this approach is to exploit the inherent sparsity of LArTPC events, allowing neural network architectures to be scalable and computationally efficient. A previous work~\cite{laura} used \emph{sparse-UResNet}--a U-Net architecture~\cite{unet} with residual connections~\cite{resnet} and 3D submanifold sparse convolutions~\cite{3dsubmfdcnn}--to do semantic segmentation of LArTPC data and achieved both higher accuracy and lower memory footprint than those of similar CNN models with dense convolutions. Following this approach, we will extend the semantic segmentation architecture and tailor the network for the task of instance segmentation. We share the same terminology used in the previous study~\cite{laura} where the architecture is defined by essentially two parameters: the number of filters in the initial convolution layer and the {\it depth}, which is the number of a spatial down-sampling operations. We shall call the block of repeated convolution and down-sampling layers an {\it encoder}, and the block of repeated convolution and up-sampling layers a {\it decoder}, as shown in Figure~\ref{fig:gm_network}. Unless stated otherwise, we construct deep learning architectures using the Pytorch base library~\cite{pytorch} with the SSCN extension~\cite{submfdcnn, 3dsubmfdcnn}. 

\subsection{Redesigning UResNet towards Instance Segmentation}

\begin{figure*}[t]
    \centering
    \includegraphics[width=0.98\textwidth]{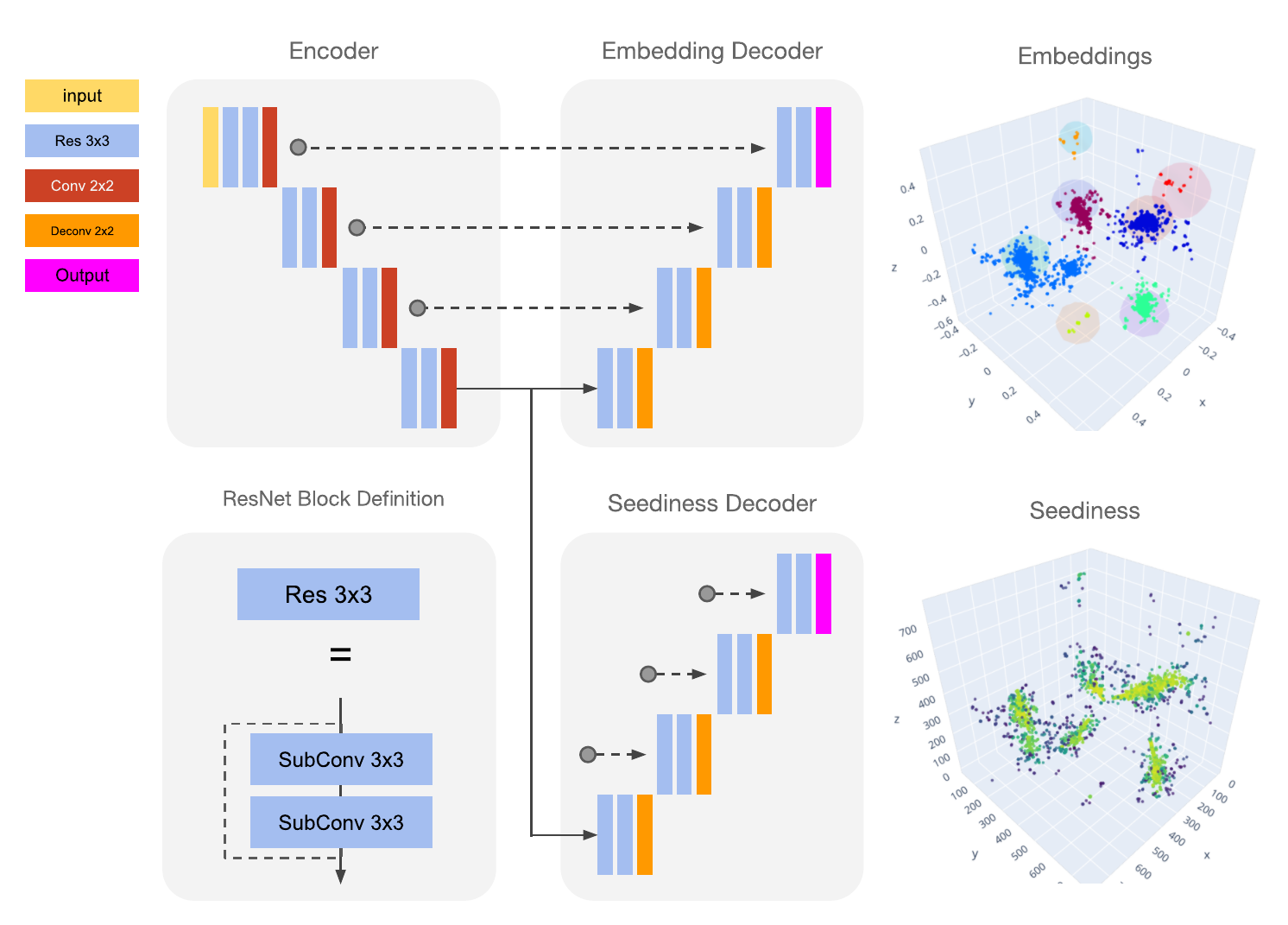}
    \caption{Network architecture for particle clustering, using sparse convolutions.}
    \label{fig:gm_network}
\end{figure*}

Our instance segmentation network closely follows a previous study in the field of Computer Vision~\cite{spatial_embeddings}. An illustration of a sample network architecture with down-sampling depth 3 is shown in Figure~\ref{fig:gm_network}. \texttt{Conv2x2} refers to 2x2 convolutions with stride 2, which function as learnable downsampling layers, and \texttt{Deconv2x2} to 2x2 learnable upsampling layers that recover the sparsity structure of the previously downsampled layer. We include batch normalization~\cite{batch_norm} and LeakyReLU~\cite{lrelu} activation layers before every convolution operations. An input 3D image with energy deposition values in each pixel is forwarded into a shared encoder. The two output data tensors, called \emph{embedding} and \emph{seediness}, are expanded to full resolution from a common spatially contracted data tensor through two separate decoders. Note that the typical U-Net skip connections from previous layers of the encoder are concatenated with intermediate data tensors in both of the decoders.

\begin{figure*}[t]
    \centering
    \includegraphics[width=\textwidth]{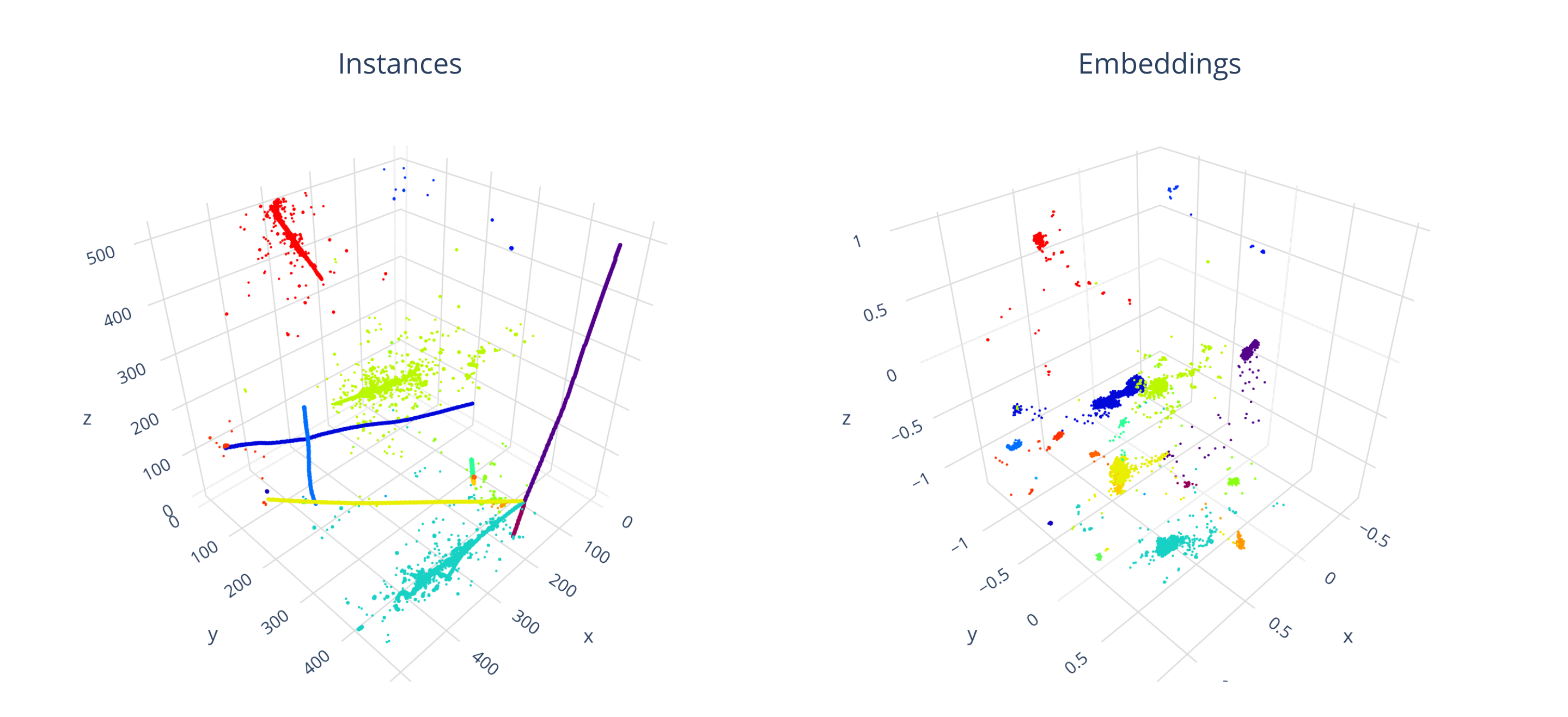}
    \caption{Example event and the learned pixel embeddings. Colors show true instance labels.}
    \label{fig:to_embeddings}
\end{figure*}

The embedding branch outputs an $N \times 4$ dimensional tensor, where the first three values correspond to the transformed coordinate in embedding space and the last value estimates the spatial extent ($\sigma_k$) of the instance $C_k$. We will first state the most general form of \emph{the embedding loss} and later consider some popular choices of binary classification losses that are available in current deep learning literature. The trained network predicts a binary mask for each instance and formulates the problems as a foreground/background binary classification problem. Consider a neural network $\tilde{f}: I \to [-1, 1]^3$, where $f_{emb}(x_i) \in E$ is defined as:
\begin{equation}
    f_{emb}(x_i) := \tilde{f}(x_i) + \tilde{x}_i, \quad \tilde{x}_i = \frac{x_i - L/2}{L/2} \in [-1,1]^3, 
\end{equation}
where $L$ is the input spatial size of the image. In other words, the vector $\tilde{x}_i \in [-1,1]^3$ is the re-scaled spatial coordinates of the pixel $x_i \in \bbZ_+^3$. Each instance $C_k$ has a centroid $\mu_k$ in embedding space defined by:
\begin{equation}
\mu_k = \frac{1}{|C_k|} \sum_{i \in C_k} f_{emb}(x_i).     
\end{equation}
An example pixel embedding generated by our network is provided in Figure~\ref{fig:to_embeddings}.

In the original paper~\cite{spatial_embeddings}, authors considered the cases where the embedding centroid $\mu_k$ is fixed. Such construction is undesirable if centroids of different clusters happen to overlap. Hence, $\mu_k$'s are always set to be trainable. Note that if we choose to minimize the euclidean distance $\norm{f_{emb}(x_i) - \mu_k}$, then $\tilde{f}$ approximates the offset vector $o_i$ from $\tilde{x}_i$ to $\mu_k$: $o_i := \tilde{x}_i - \mu_k$. Now define $p: \bbR^3 \times \bbR^3 \times \bbR^1: \to [0,1]$ as:
\begin{equation}
    p_{ik} \equiv p(x_i; \mu_k, \sigma_k) = \exp{\left[-\frac{\norm{f_{emb}(x_i) - \mu_k}^2}{2 \sigma_k^2} \right]},
\end{equation}
where $x_i \in \bbR^3$, $\mu_k \in \bbR^3$, and $\sigma_k \in \bbR^1_+$. 
This defines a spherical gaussian kernel, centered at $\mu_k$ with bandwidth $\sigma_k$. We interpret this value as the score value for a pixel $x_i$ to belong to cluster $k$ with embedding size $\sigma_k$. For each instance $\{C_k\}_{k=1}^K$, the kernel $p(\cdot; \mu_k, \sigma_k)$ defines a similarity kernel that produces a score map for all pixels within the same semantic class. We can then evaluate the quality of our pixel embeddings by measuring, for example, the binary cross entropy (BCE) score between the score map and the instance labels:
\begin{equation}
    \mathcal{L}_{BCE}^{(k)} = -\frac{1}{N} \sum_{i = 1}^N [y_{ik} \log{(p_{ik})} + (1 - y_{ik}) \log{(1 - p_{ik})}], 
\end{equation}
where $y_{ik} = 1$ if $i \in C_k$ and 0 otherwise. The embedding generating network is trained by minimizing $\mathcal{L}_{BCE}$, so that the embeddings $f(x_i)$ arrange in such a way that $p_{ik} \approx 1$ if $f(x_i)$ belongs to $C_k$ and $p_{ik} \approx 0$ otherwise. The \emph{mask loss} for a given data tensor is then defined as the $\mathcal{L}_{BCE}$ averaged over the instances:
\begin{equation}
\mathcal{L}_{mask} = \frac{1}{K} \sum_{k = 1}^K \mathcal{L}_{BCE}^{(k)}. 
\label{mask_loss}
\end{equation}

The choice of the embedding loss function is not limited to $\mathcal{L}_{BCE}$. Some popular loss function in semantic and instance segmentation include soft dice loss~\cite{dice_loss}, Lov\'asz hinge and softmax~\cite{lovasz-hinge}, and focal loss~\cite{focal_loss}. Each of these functions may replace the binary cross entropy loss of equation \eqref{mask_loss} to define a new loss function for training the embedding network. In the original study~\cite{spatial_embeddings}, the authors used the Lov\'asz hinge loss (also known as smooth intersection-over-union loss) to directly maximize the intersection-over-union of each predicted instance mask. As such, the Lov\'asz hinge loss is used for the results presented in this study.

 To make the $\sigma$ values consistent throughout pixels in the same instance, we introduce a smoothing loss that constrains the margin values $\sigma_i$ for $i \in C_k$. The \emph{smoothing loss} is defined as:
\begin{equation}
    \mathcal{L}_{smoothing} := \frac{1}{|C_k|} \sum_{i = 1}^{N_k} \norm{\sigma_i - \tilde{\sigma}_k}, \quad \tilde{\sigma}_k := \frac{1}{|C_k|} \sum_{i = 1}^{N_k} \sigma_i,
\end{equation}
where $\{\sigma_1, ..., \sigma_{N_k}\}$ are predicted margin values of pixels that belong to the same cluster $C_k$. Then $\sigma_k$ is the average predicted margin value of cluster $C_k$. Note that $\tilde{\sigma}_k$ indicates that the smoothing loss is evaluated with $\sigma_k$ detached from the back-propagation graph. 

From practice, we observed that including an inter-cluster loss that penalizes small distances between two cluster centroids further improves separation between the cluster embeddings. From \cite{disc}, the \emph{inter-cluster} loss between two cluster centroids $\mu_i$ and $\mu_j$ are defined as:
\begin{equation}
    \mathcal{L}_{inter} = \frac{1}{K(K-1)} \sum_{i < j} [2\delta_d - \norm{\mu_i - \mu_j}]_+^2,
\end{equation}
where $\delta_d$ is a inter-cluster margin parameter.

The output data tensor of the embedding branch is then trained on the combined loss 
\begin{equation}
\mathcal{L}_{emb} = \mathcal{L}_{mask} + \mathcal{L}_{smoothing} + \mathcal{L}_{inter}.
\end{equation}

During inference, we cannot compute $\mu_k$ and $\sigma_k$ as we do not have access to true labels. Hence, we train the second decoder to generate a seediness map, which quantifies how likely a given pixel is to be an instance centroid in embedding space. Since our similarity kernel $p(x_i; \mu_k, \sigma_k)$ already gives this measure, we simply train the seediness branch output to predict the values $p_{ik}$ for pixel $i$ belonging to instance $k$. 

The \emph{seediness loss} is defined as:
\begin{equation}
    \mathcal{L}_{seed}^{(k)} := \frac{1}{|C_k|} \sum_{i \in C_k} \norm{s_i - p(x_i; \mu_k, \sigma_k)}^2,
\end{equation}
where $s_i \in [0,1]$ is an estimate of how likely a given pixel $x_i$ is close (in embedding space) to the true centroid $\mu_k$.
The output of the seediness branch $s_i \in (0,1)$ is constrained with a Sigmoid activation layer and detached $p(x_i ; \mu_k, \sigma_k)$ from the computation graph when evaluating $\mathcal{L}_{seed}$. Again, the full seediness loss is obtained by averaging over different instances: 
\begin{equation}
\mathcal{L}_{seed} = \frac{1}{K} \sum_{k = 1}^K \mathcal{L}^{(k)}_{seed}.
\end{equation}

\subsection{Training and Inference}
We constrain the values of the offset vectors and the margin values to $o_i \in [-1, 1]$ and $\sigma_i \in [0, 2]$ by passing the features through Tanh and Sigmoid activation layers (with a constant multiplicative factor of 2). 

The combined loss is given as
\begin{equation}
    \mathcal{L}_{total} =  \gamma_1 \mathcal{L}_{seed} + \gamma_2 \mathcal{L}_{emb}.
\end{equation}
During training, we compute the true instance embedding centroid $\mu_k$ and margin $\sigma_k$ using ground-truth semantic and instance labels. The contributions from each cluster are averaged within each class, event, and minibatch. By default, we use the weights $\gamma_1 = 1$ and $\gamma_2 = 1$.

Inference employs an iterative greedy search procedure for detecting instances following the original paper~\cite{spatial_embeddings}. For each semantic class, we obtain the predicted seediness values for each pixel in the current semantic class. After sorting the seediness values in decreasing order, we begin with the highest seediness $s_k$ and query the corresponding embedding vector $f(x_i)$ and the margin value $\sigma_k$. This defines a gaussian kernel $p( \ \cdot \ ; {\mu}_k, \sigma_k)$, from which we compute the probability value $p_{ik}$ for every embedding vector $f(x_i)$ belonging to the current semantic class. We then assign an instance label to all $f(x_i)$ that have $p_{ik} > p_0$, where $p_0 \in [0,1]$. Accepting the probability interpretation of $p(\mathbf{f}; \hat{\mu}_k, \sigma_k)$, we may choose $p_0 = 0.5$, for example. We then repeat the same process beginning with the highest seediness point that was not given an instance label, until either 1) we cluster all points in a given semantic class or 2) the remaining pixels have seediness values less than a predetermined minimum value $s_0$. After this procedure, we obtain a list of cluster centroids and the corresponding probability maps. The instance labels are given for each pixel $x_i$ by assigning it to the cluster with the highest $p_{ik}$, and the same procedure is repeated for all semantic classes in a given image. 

\section{Experiment}

\subsection{Training Details}
\begin{table}[t]
\caption{Summary of testing dataset. $N_{clusters}$ is the total number of instances of the specified semantic type, $N_{events}$ is the number of events containing the specified semantic type, and $\bar{n}_{clusters}$ is the average number of instances in one event. For example, we have an average of $5 \pm 3$ MIP tracks per event in our testing dataset.}

\begin{tabular}{ccccc}  
\toprule
\multicolumn{5}{c}{Test Set} \\
\cmidrule(r){1-5}
& {$N_{clusters}$} & {$N_{events}$} & {$\lfloor \bar{n}_{clusters} \rfloor$} & {Pixel Counts}\\
\midrule
HIP & {353,555} & {17,460} & {$4 \pm 3$} & {32.1M} \\
MIP & {572,794} & {19,278} & {$5 \pm 3$} & {122M} \\
Shower & {1,528,505} & {19,476} & {$8 \pm 4$} & {193M} \\
Delta & {329,689} & {13,861} & {$4 \pm 2$} & {3.94M} \\
Michel & {38,957} & {6,122} & {$2 \pm 1$} & {2.29M} \\
\bottomrule
\end{tabular}
\label{tab:stats}
\end{table}
We train sparse-UResNet with an initial convolution layer with a filter count of 32 and depth 6, using the Adam optimizer~\cite{adam} with an initial learning rate of 0.005. Following the previous work, the number of filters at each depth is increased linearly as the depth increases in the architecture. The training set consists of 80,000 LArTPC simulation sample with 512px resolution (512x512x512 3D images). The total number of clusters and pixel counts per semantic type in the testing dataset are summarized in Table~\ref{tab:stats}. Training each model is done via a two-step process. First, we optimize the embedding branch of the network by training on the embedding loss $\mathcal{L}_{emb}$ while freezing the seediness branch of the network. This is to ensure that the embeddings are well separated enough with respect to each clusters so that the seediness values $p_{ik}$ give sensible score values during training. The embedding branch is first trained for 50,000 iterations with batch size 32, and the full network including the seediness branch is then trained for additional 100,000 iterations until convergence.

\subsection{Choice of Clustering Performance Metrics}

Our Monte-Carlo simulation sample provides true particle instance labels for each pixel in an input image. To evaluate clustering accuracy, we use the \emph{adjusted Rand index} (ARI)~\cite{ari} and generalized purity and efficiency scores, which we define as follows:

Let $Y^* = \{S_1^*, S_2^*, ..., S_n^*\}$ denote the true partitioning of $\Omega$ and let $\tilde{Y} = \{\tilde{S}_1, \tilde{S}_2, ..., \tilde{S}_m\}$ be the predicted partitioning of $\Omega$. Let $c_{ij}$ denote the entries of the contingency matrix $\mathcal{C}(Y^*, \tilde{Y})$, where $c_{ij} = |S^*_i \cap \tilde{S}_j|$. The \emph{purity} $\mathcal{P}$ of $\tilde{Y}$ with respect to $Y^*$ is defined as
\begin{equation}
\mathcal{P}(Y^*, \tilde{Y}) = \frac{1}{m} \sum_{j = 1}^m \frac{\displaystyle \max_{i = 1, ..., n} c_{ij}}{|\tilde{S}_j|}. 
\end{equation}
Likewise, the \emph{efficiency} $\mathcal{E}$ is defined as
\begin{equation}
\mathcal{E}(Y^*, \tilde{Y}) = \frac{1}{n} \sum_{i = 1}^n \frac{\displaystyle \max_{j = 1, ..., m} c_{ij}}{|S^*_i|}. 
\end{equation}

Let $n_{pred}$ be the number of instances predicted by the network and let $n_{true}$ be the true number of clusters. It is easy to show from the analytical formula~\cite{ari} that the ARI metric evaluates to zero if $n_{true} \neq n_{pred}$ and either $n_{true} = 1$ or $n_{pred} = 1$. More precisely, if a given instance consisting of 100 pixels are separated into two clusters of 99 pixels and 1 outlier pixel, the ARI metric will evaluate to zero even if the predicted labels does not deviate much from true labels. Indeed, we observed that there are many events with ARI value being exactly zero due to this property of the ARI metric and are not much informative on the potential shortcomings of the proposed method. Hence, in searching for common mistakes and evaluating performance, we search among those events with both true and predicted number of clusters higher than one.

\subsection{Inference Details}
\begin{table}[t]
\centering
\caption{Optimal thresholding parameters $s_0, p_0$ obtained via grid search over 100 validation set events.}
\begin{tabular}{ccc}  
\toprule
\multicolumn{3}{c}{Optimized Thresholding Values} \\
\cmidrule(r){1-3}
& {$s_0$} & {$p_0$}  \\
\midrule
HIP & 0.65 & 0.29  \\
MIP & 0.5 & 0.087 \\
Shower & 0.25 & 0.036\\
Delta & 0.85 & 0.27 \\
Michel & 0.0 & 0.087 \\
\bottomrule
\end{tabular}
\label{table2}
\end{table}
We perform inference on a validation set of 19,900 events, with 100 events separated for hyperparameter optimization. The performance of the label generating step is conditioned on two hyperparameters: the thresholding values $s_0, p_0$. For each semantic class, we performed a grid search over the values $s_0 \in [0, 1]$ and $p_0 \in [0, 0.6]$ using a validation set of 100 events, separate from both the training and the test set. The optimized values are provided in Table~\ref{table2}.

Lastly, during evaluation on the test set we mask out all true clusters with less than 10 pixels. Such small pixel instances arise when a given instance is separated into different semantic classes. These most often happen at the interface between different pixel types, such as a muon decaying into a Michel electron. Also, since 10 pixel clusters are unlikely to affect any physics analysis targets, we shall omit these instances when computing accuracy metrics over events in the test set.

%
%
%
%

\section{Results}
\subsection{Predicted Clusters and Margins}
\begin{figure*}[t]
\centering
    \includegraphics[width=0.85\textwidth]{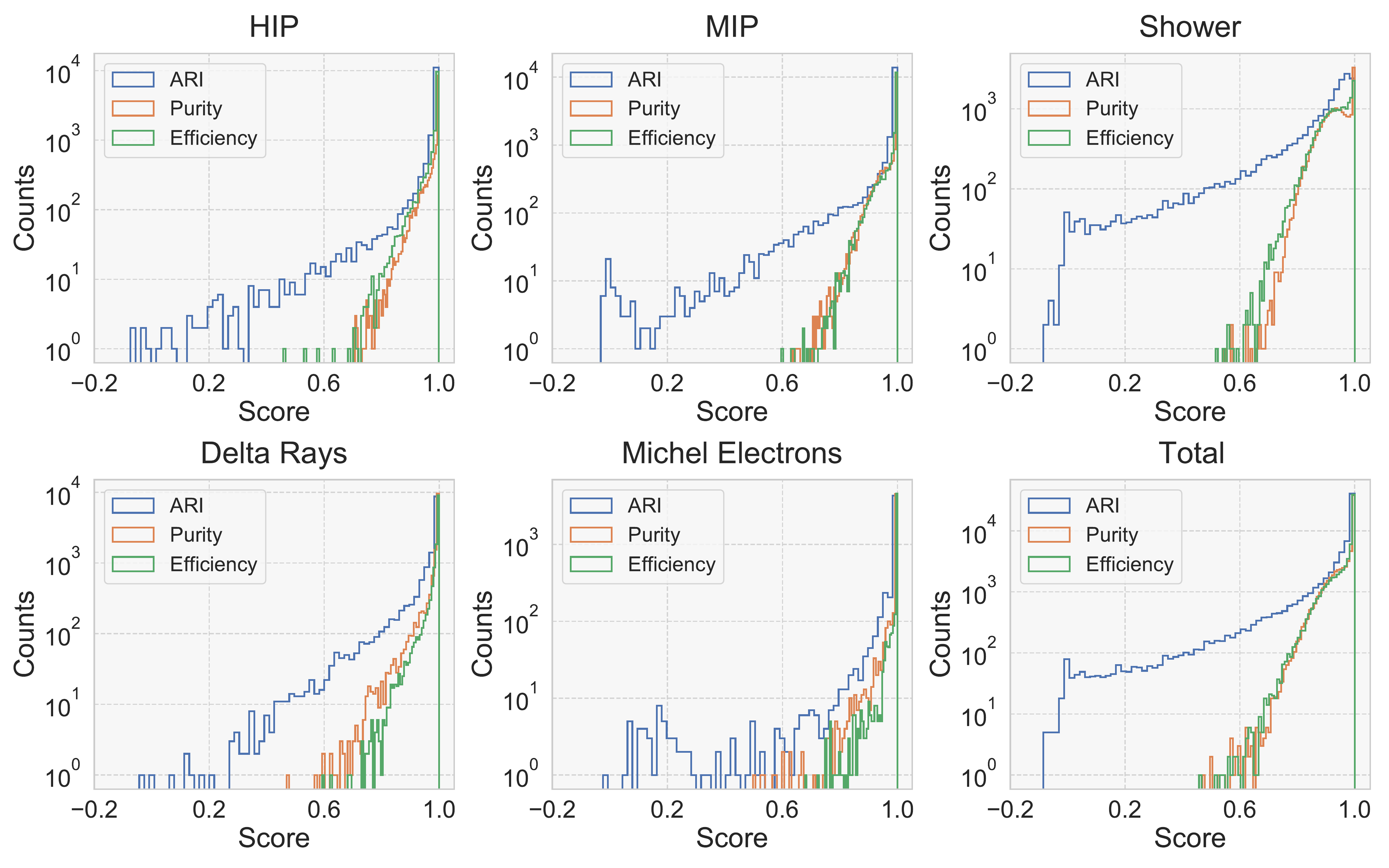}
    \caption{Plots showing the distribution of accuracy metrics for a selected model. The $y$-axis is in log scale.}
    \label{fig:acc_baseline}

    \includegraphics[width=0.86\textwidth]{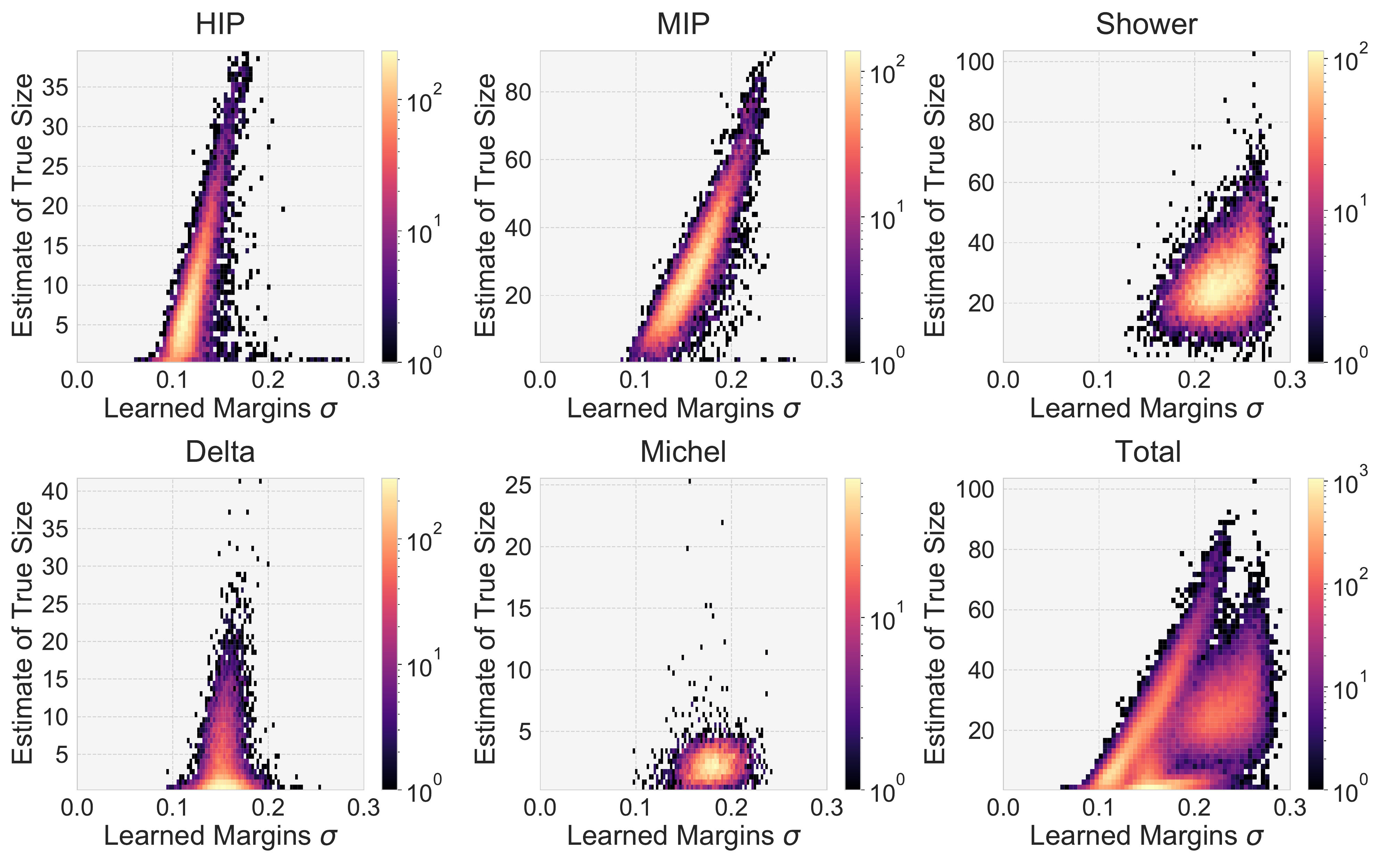}
    \caption{Learned margin values $\sigma$, with respect to estimate of true instance size computed by the standard deviation of the pixel distances from the instance center in image coordinates.}
    \label{fig:margins}
\end{figure*}


\begin{figure*}[tbh]
    \includegraphics[width=1.00\textwidth]{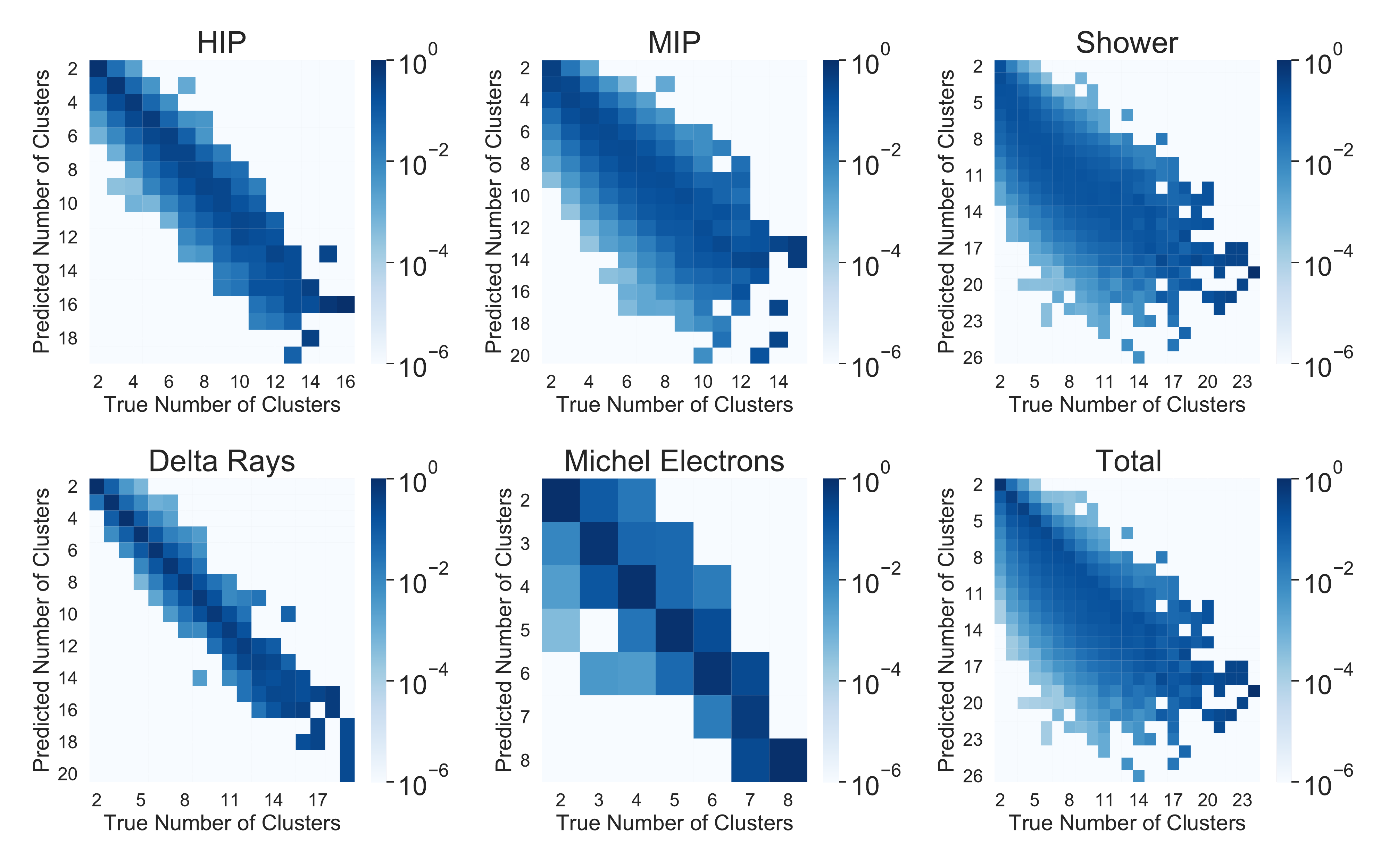}
    \caption{Normalized confusion matrix of the predicted number of clusters.}
    \label{fig:counts}
\end{figure*}

\begin{figure*}[tbh]
    \includegraphics[width=1.00\textwidth]{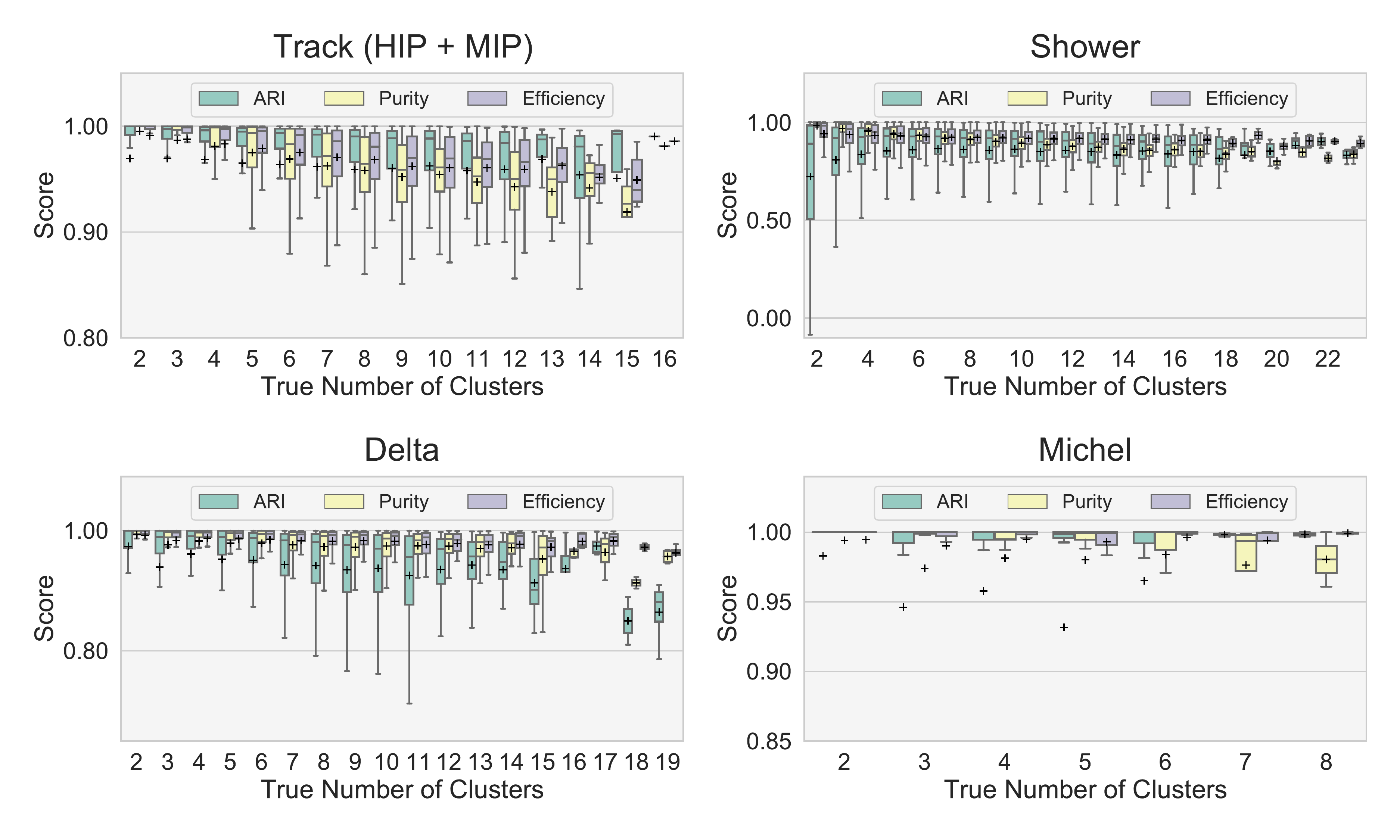}
    \caption{Performance metric behavior across different true number of particles.}
    \label{fig:num_clusters}
\end{figure*}

\begin{figure*}[tbh]
\vspace{0.1in}
\centering
  \includegraphics[width=1.00\textwidth]{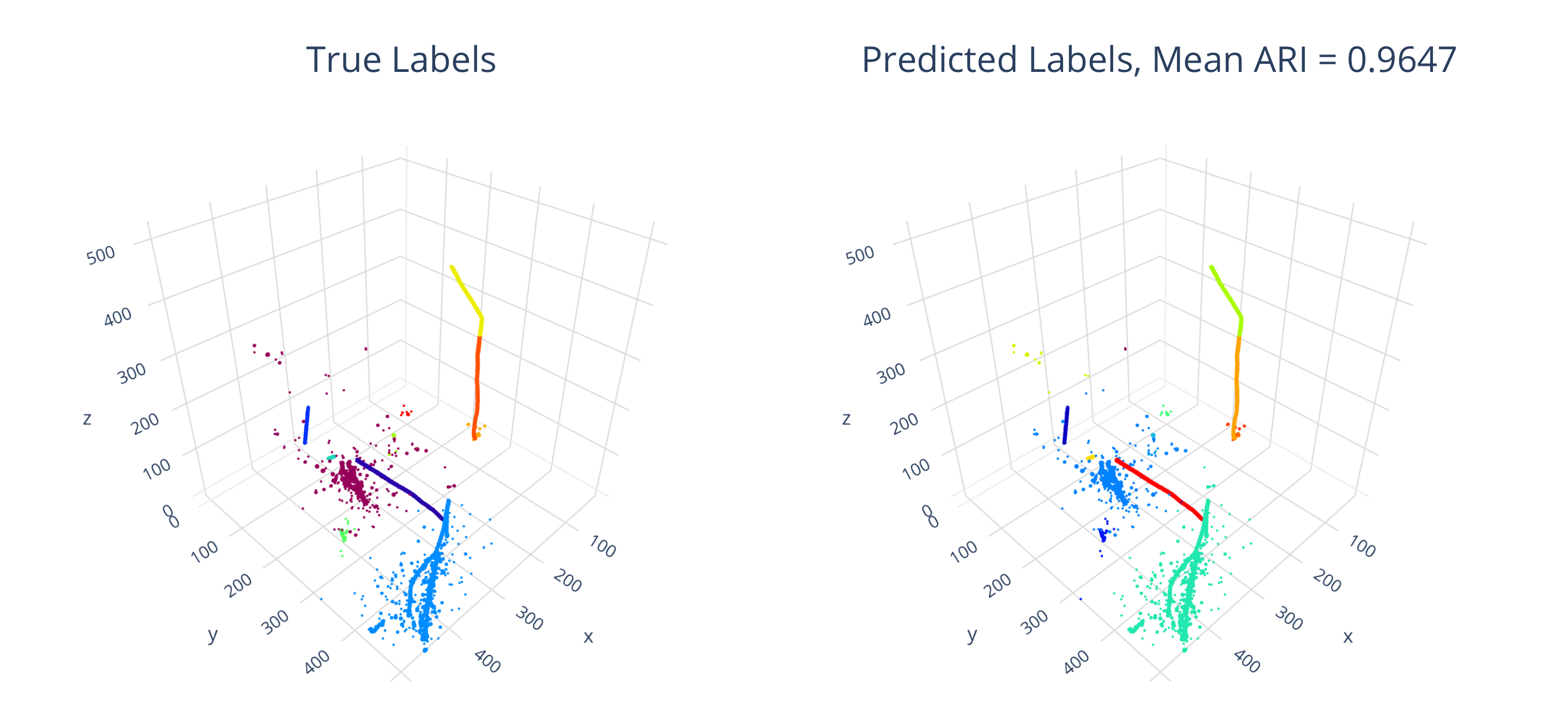}\\
  
  \vspace{0.2in}
  \includegraphics[width=1.00\textwidth]{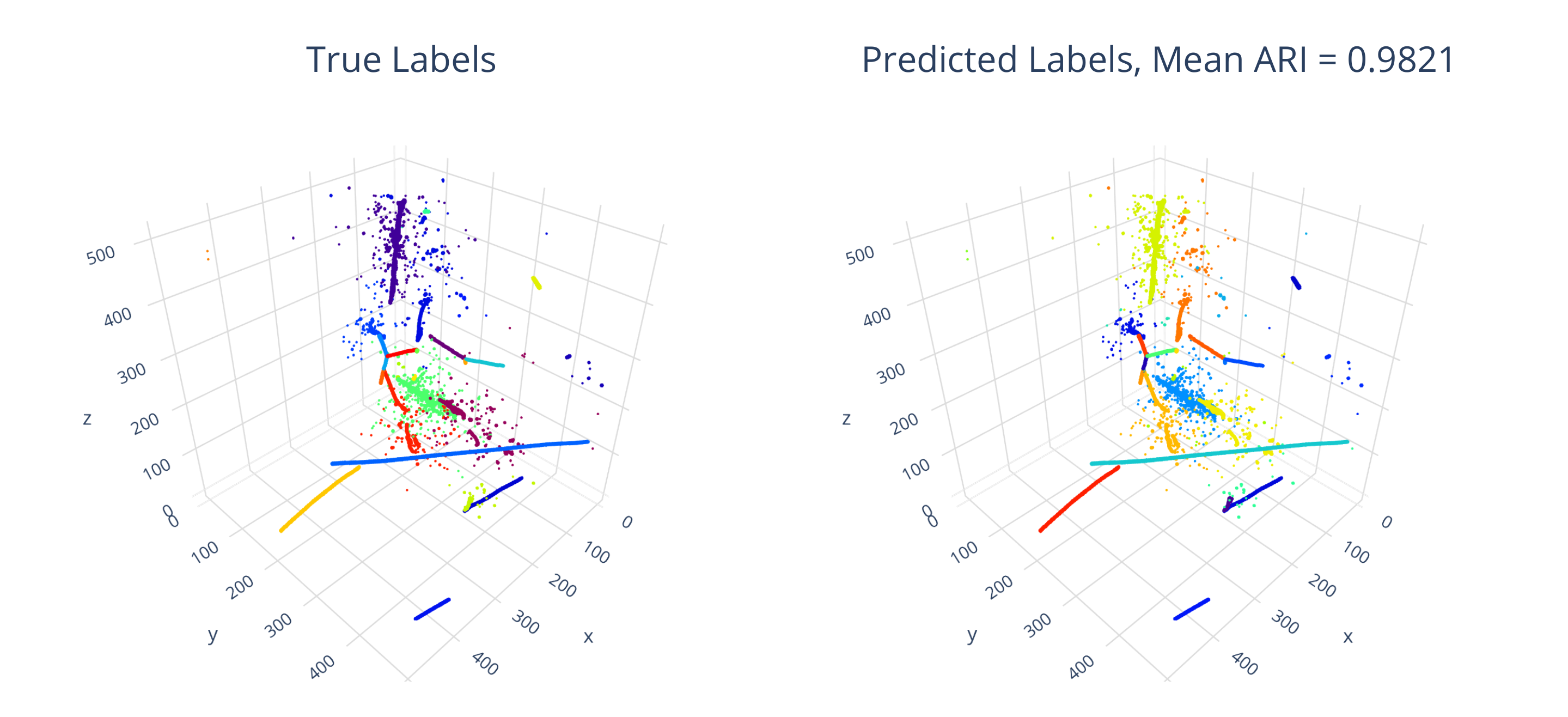}
\caption{Comparison of true (left column) and predicted (right column) instance labels from randomly selected two images using the best model. Each particle instance is shown in a distinct color. Colors are randomly assigned in each event display and there is no matching of colors between the true labels and the prediction.}
\label{fig:samples}
\end{figure*}


\begin{table}[t]
\caption{Mean accuracy metrics within each semantic class, computed over 19,900 samples of the test set.}
\begin{tabular}{cccc}  
\toprule
\multicolumn{4}{c}{Test Set} \\
\cmidrule(r){2-4}
& {Purity} & {Efficiency} & {ARI} \\
\midrule
HIP & {0.9824} & {0.9843} & {0.9743} \\
MIP & {0.9783} & {0.9809} & {0.9612} \\
Shower & {0.9317} & {0.9275} & {0.8408} \\
Delta & {0.9832} & {0.9873} & {0.9567} \\
Michel & {0.9906} & {0.9941} & {0.9746} \\
\bottomrule
\end{tabular}
\label{tab:table3}
\end{table}

We first evaluate the overall clustering performance of the trained model using the accuracy metrics defined in section 2. The distributions of different accuracy metrics are visualized in Figure~\ref{fig:acc_baseline}. We see in general that the network succeeds in clustering most of the instances with high accuracy metric scores, with the exception of EM showers, where we observe a longer tail towards lower ARI values. This aligns with our intuition that EM showers, due to their complex topology of scattered energy depositions, are more difficult to cluster into separate instances than other classes such as HIP/MIP tracks. The mean value of the performance metrics are shown in Table~\ref{tab:table3}. In particular, our algorithm achieved mean efficiency/purity/ARI of 0.98/0.98/0.97 for track-like particles (i.e. MIP and HIP particles), and 0.93/0.93/0.84 for EM showers.


Another aspect of the model is the correlation between the spatial extent of an instance in image coordinate space and the learned margins $\sigma_i$ in embedding space. This correlation was also observed in the original paper~\cite{spatial_embeddings}. Figure~\ref{fig:margins} show a 2D histogram of the correlation between the prediction Gaussian width (margin) values $\sigma$ and the true spatial extent of the instance, which is defined as the standard deviation of the distance from the instance center in image coordinate space. The correlation is strongest in the case for tracks (HIP and MIP), while for deltas and Michel electrons the predicted margins do not seem to scale with increasing spatial extent.

Next, we analyze the effect of the true number of instances on clustering accuracy. Figure~\ref{fig:counts} show the correlation between the number of true particle instances and the number of predicted instances. Figure~\ref{fig:num_clusters} shows the extent to which the number of instances affect the network performance on the test set, where the boxes represent the interquartile range (IQR) and the whiskers represent $\pm$1.5 IQR from the median. We observe evidence of slight performance degradation towards higher number of clusters.

\begin{figure*}[t]
\centering
  \includegraphics[width=0.88\textwidth]{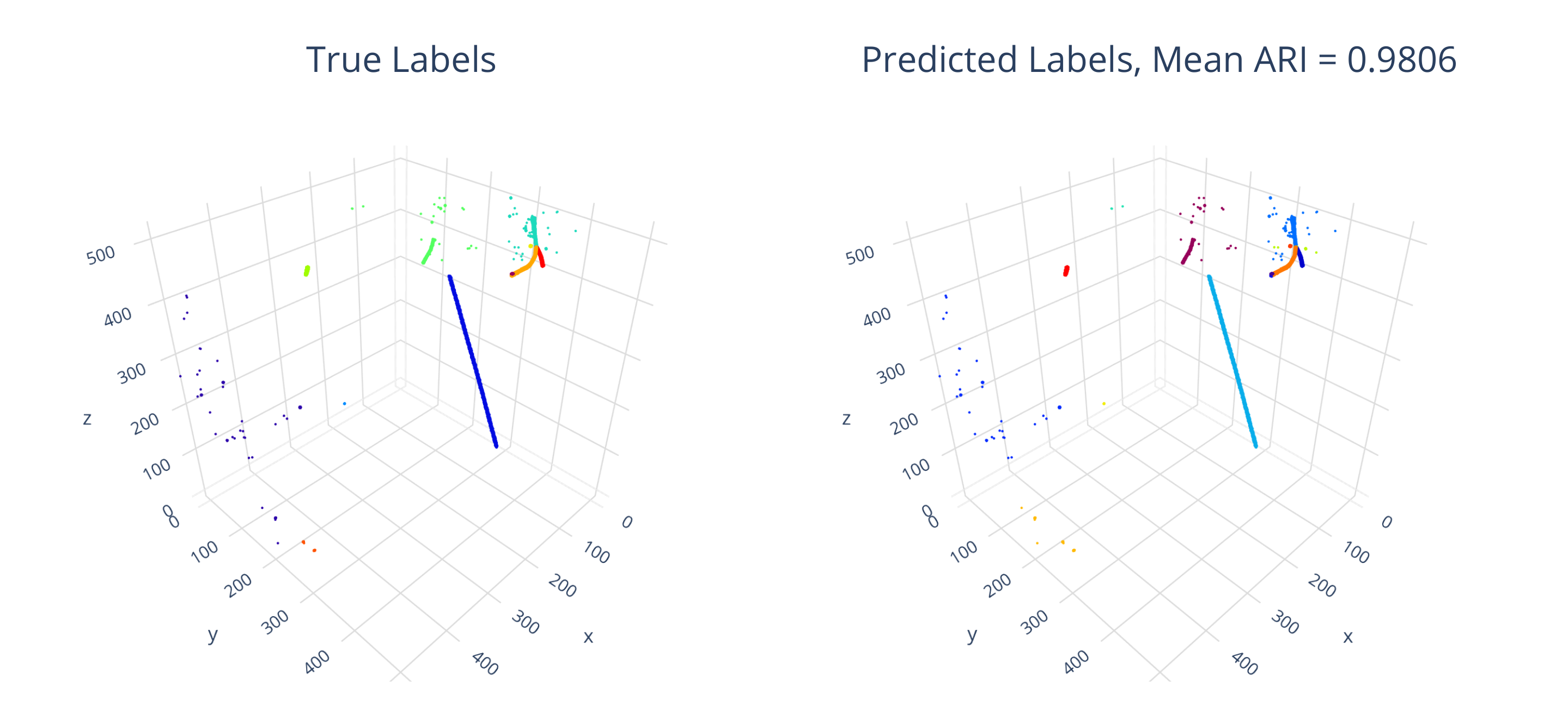}\\
  \includegraphics[width=0.88\textwidth]{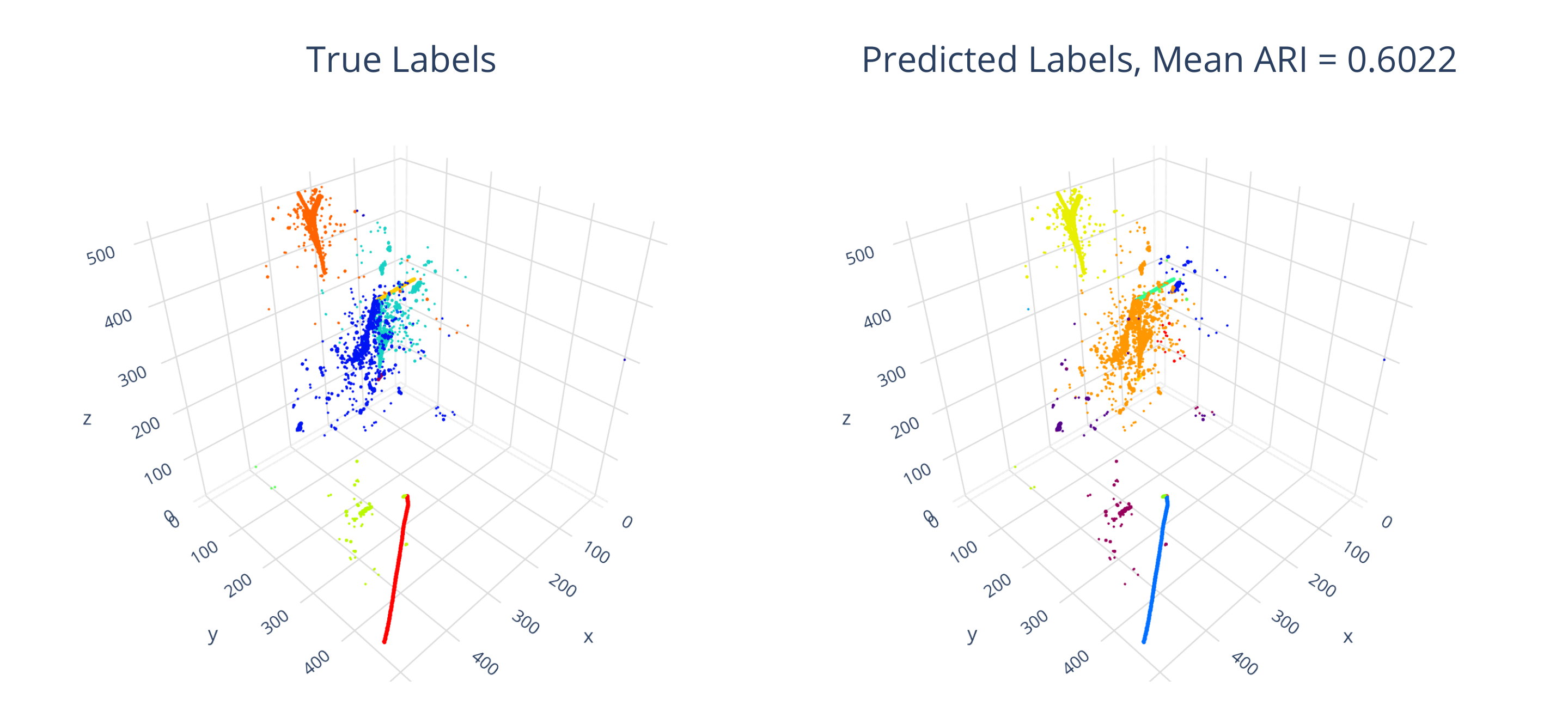}\\
  \includegraphics[width=0.88\textwidth]{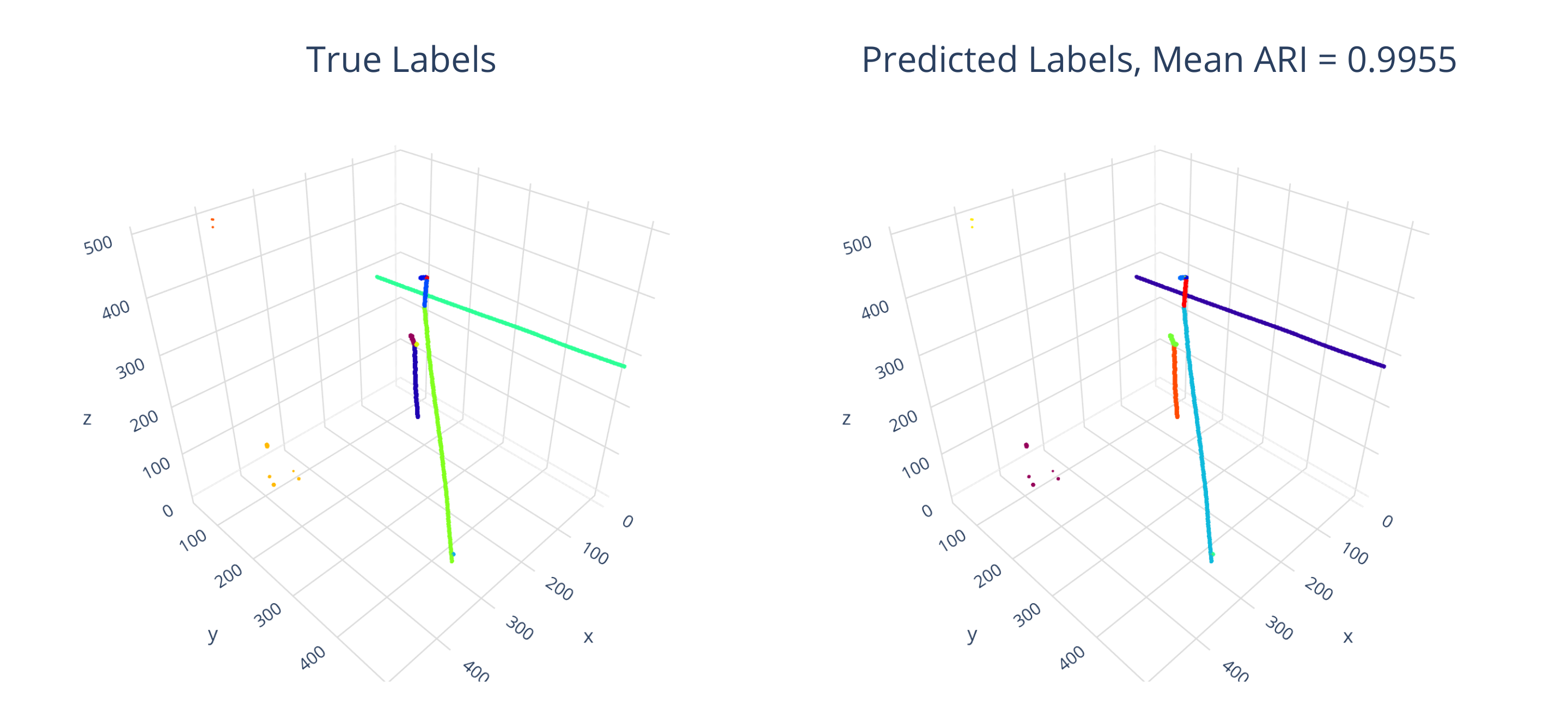}
\caption{Additional examples: comparison of true (left column) and predicted (right column) instance labels from randomly selected three images using the best model. Each particle instance is shown in a distinct color. Colors are randomly assigned in each event display and there is no matching of colors between the true labels and the prediction. }
\label{fig:more_samples}
\end{figure*}

A randomly selected collection of network predictions compared with true labels are presented in Figure~\ref{fig:samples}. Although it is possible to see several mistakes, in other cases the proposed method is successful in segmenting different particle instances, especially in separating track instances that are contiguous in image space. More examples can be found in Figure~\ref{fig:more_samples}.

\subsection{Error Analysis}




Although the proposed method is overall successful in clustering particles, in some cases the network and the post-processing algorithm do not generate labels that closely match the true instance labels. Here we list general patterns for the most common mistakes, which are illustrated with event displays in the appendix. The 3D images in the top row show original image pixels, while the bottom row display pixel embeddings. 

\subsubsection{Low pixel count clusters due to overlapping pixels}
The semantic and instance labels provided in PILArNet~\cite{PublicSample} are defined in such a way that a single pixel can belong to multiple particle instances while its true semantic type is uniquely defined. This happens, for example, at the interface between a muon and a Michel electron. Since clustering is done after masking the event by semantic labels, these overlapping regions appear as separate clusters of pixels with generally low pixel counts. 
As such, most of these clusters do not represent true particle instances but rather an artifact of clustering particles within a fixed semantic class. The aforementioned 10 pixel cut does not completely resolve this issue, and the remaining ill-defined clusters tend to lower the average accuracy metric across all classes. 

\subsubsection{Long tracks separated into multiple segments}
\begin{figure*}[t]
    \centering
    \includegraphics[width=0.97\textwidth]{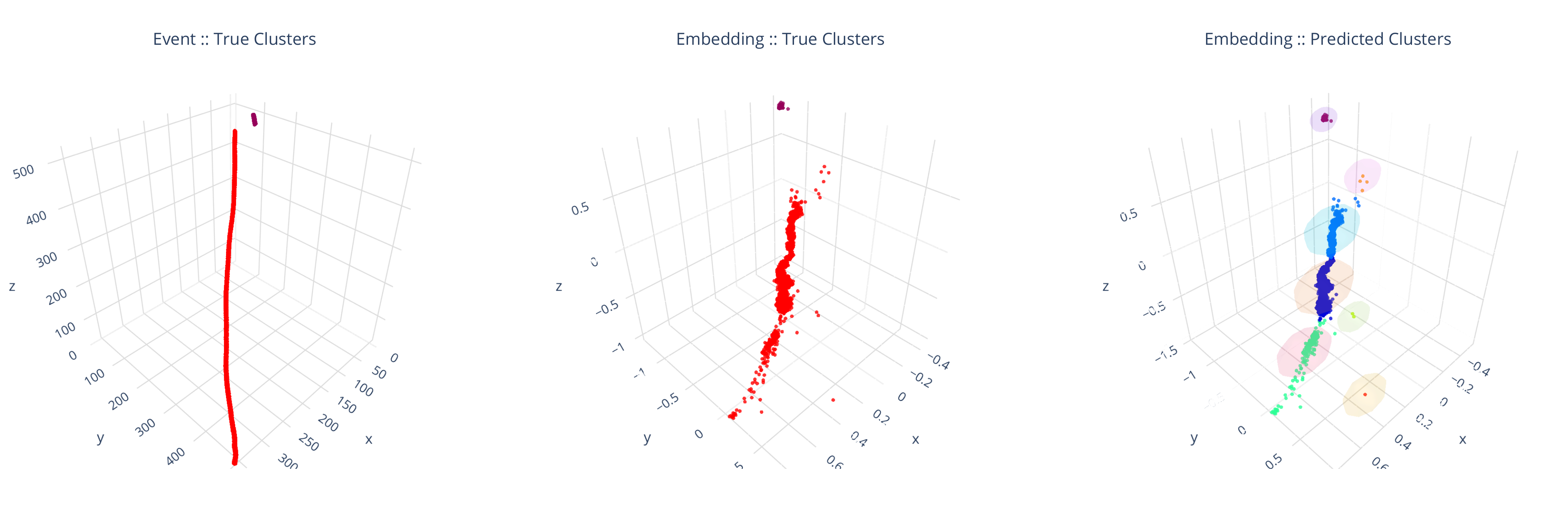}
    \caption{A long MIP track is segmented to multiple clusters due to an unsatisfactory pixel embedding.  From the left, trajectories with true instance labels shown in colors, the same true labels shown in the embedding space as a result of the network, and predicted instance labels shown in colors as a result of post-processing.}
    \label{fig:mistake_long_track}
\end{figure*}

 Common problem among tracks that roughly span the whole image is that the embedding branch fails to accumulate pixel embeddings in a spherical cluster consistent with post-processing as shown in Figure~\ref{fig:mistake_long_track}. By examining the same issue with different models, we observed that the problem is more severe for models with lower number of parameters. 

\subsubsection{Co-linear and adjacent tracks}
\begin{figure*}[t]
    \centering
    \includegraphics[width=0.97\textwidth]{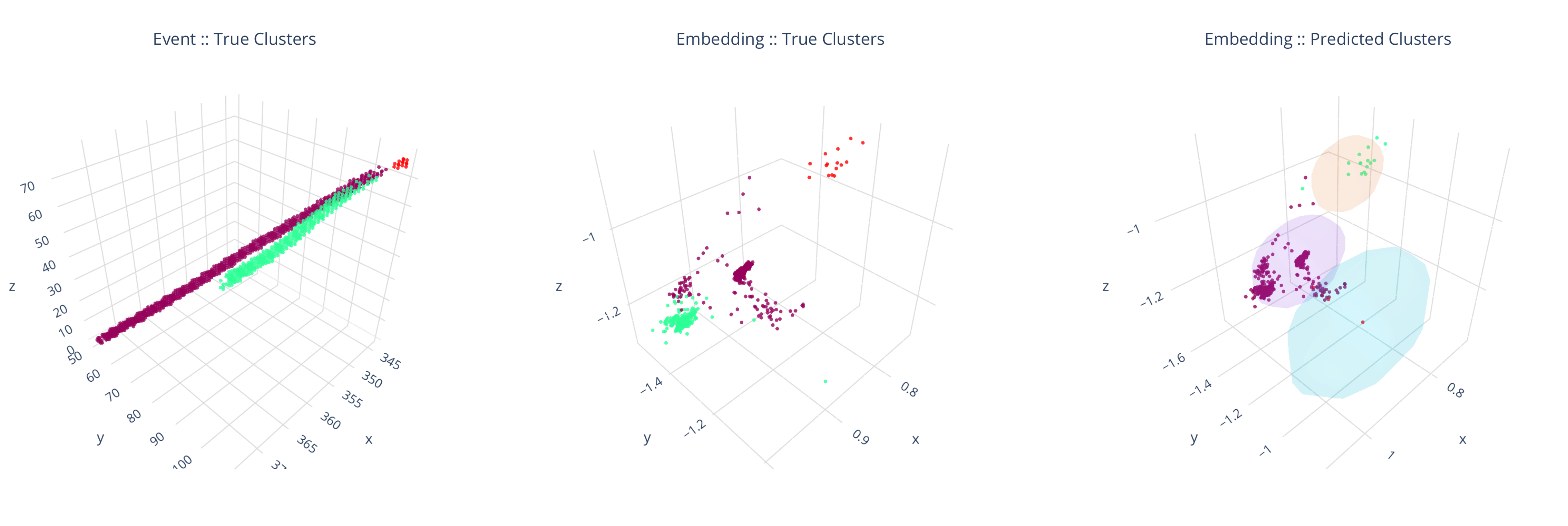}
    \caption{Nearly co-linear track instances grouped as single particle.  From the left, trajectories with true instance labels shown in colors, the same true labels shown in the embedding space as a result of the network, and predicted instance labels shown in colors as a result of post-processing.}
    \label{fig:mistake_colinear_track}
\end{figure*}

When two track instances emanate from the same vertex with almost co-linear direction vectors, then the network often fails to generate a suitable embedding that distinguishes the two track instances as shown in Figure~\ref{fig:mistake_colinear_track}

\subsubsection{Deficiencies in post-processing}
\begin{figure*}[t]
    \centering
    \includegraphics[width=0.97\textwidth]{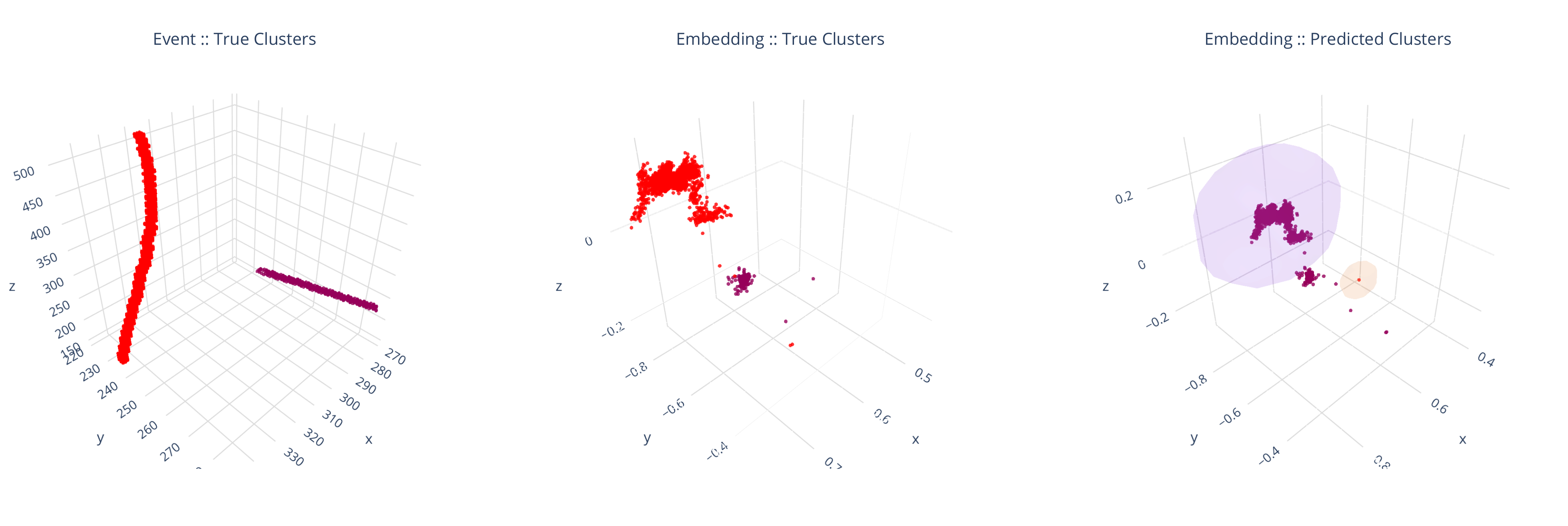}
    \caption{A post-processing stage deficiency that groups two distinct embeddings into one. From the left, trajectories with true instance labels shown in colors, the same true labels shown in the embedding space as a result of the network, and predicted instance labels shown in colors as a result of post-processing.}
    \label{fig:mistake_post_processing}
\end{figure*}

Sometimes, the network will generate a reasonable embedding that separates two instances but erroneously merge them into one during post-processing. An example is given in Figure~\ref{fig:mistake_post_processing}. Given a clear separation shown in the example, there is room for improving the post-processing procedure. While that may be pursued in the future, exploration of more complicated post-processing schemes is outside the scope of this paper.

\subsubsection{Overlapping showers}
\begin{figure*}[t]
    \centering
    \includegraphics[width=0.98\textwidth]{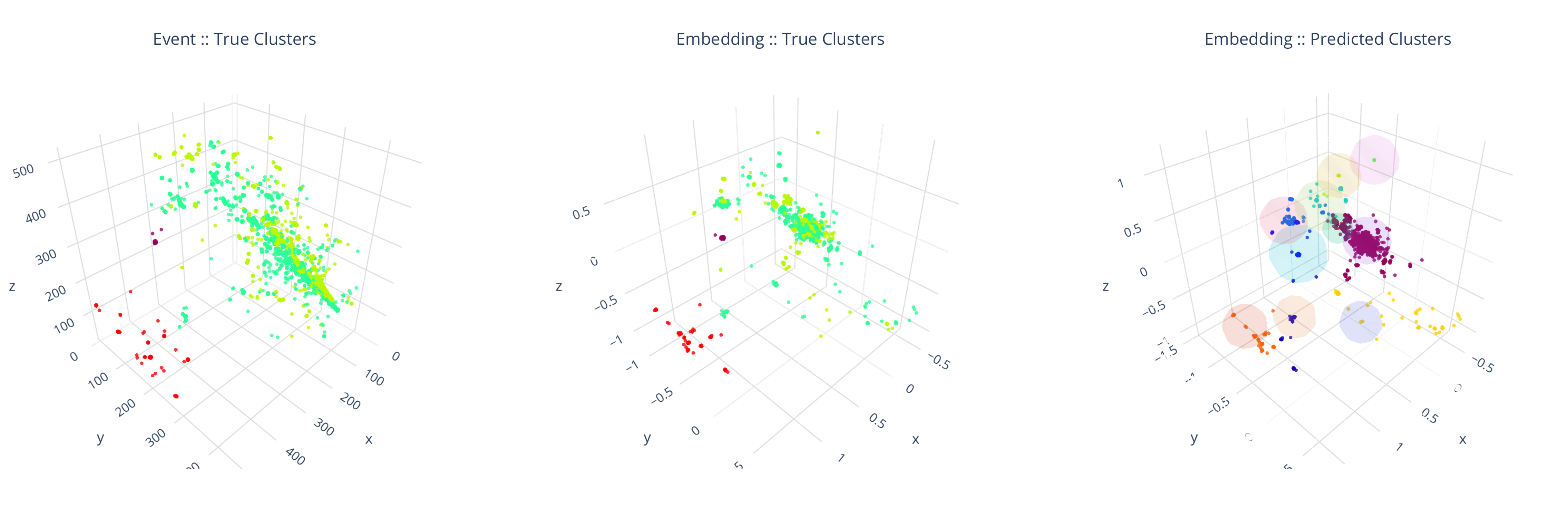}
    \caption{Overlapping shower instances grouped into one particle.  From the left, trajectories with true instance labels shown in colors, the same true labels shown in the embedding space as a result of the network, and predicted instance labels shown in colors as a result of post-processing.}
    \label{fig:mistake_overlapping_showers}
\end{figure*}

Another complication in shower clustering arises when two shower instances intersect, or if the primary ionization is nearly co-linear. Under such shower geometry, the network tends to group all pixels to the same cluster, as we observe in Figure~\ref{fig:mistake_overlapping_showers}. 

\subsubsection{Low energy depositions in EM showers}
\begin{figure*}[t]
    \centering
    \includegraphics[width=0.98\textwidth]{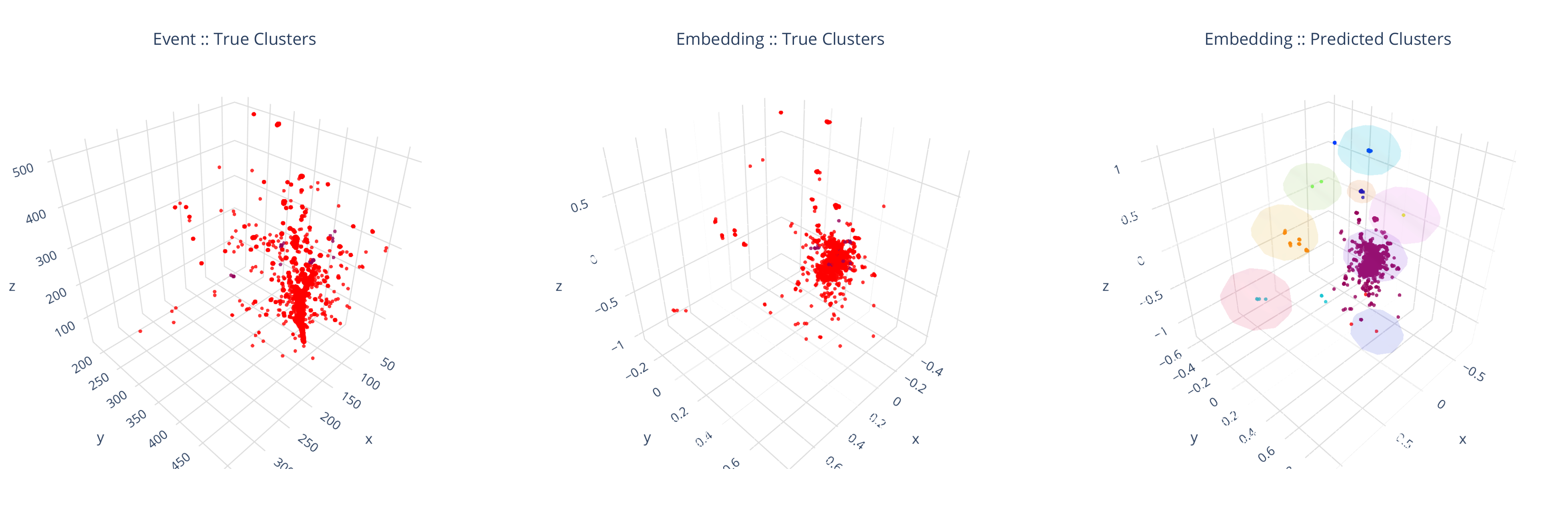}
    \caption{A shower instance where low energy depositions are incorrectly clustered as particle instances.  From the left, trajectories with true instance labels shown in colors, the same true labels shown in the embedding space as a result of the network, and predicted instance labels shown in colors as a result of post-processing.}
    \label{fig:mistake_shower_lowE}
\end{figure*}

EM shower instances contain low energy depositions that are scattered around a primary ionization trace. For shower clustering, the network often predicts a large cluster around the primary ionization of a shower and cluster the remaining low energy depositions as separate instances. Moreover, in some cases, the true shower instance does not have a clearly visible primary ionization track, so that a given shower instance resembles a spatially disperse cloud of low energy depositions. Assigning low energy depositions to the correct shower instance is a naturally difficult (if not ill-defined) task, and some of the performance degradation for shower clustering could be explained by these types of mistakes. This group of mistakes could be best represented by Figure~\ref{fig:mistake_shower_lowE}.

\subsection{Computational Resource Usage}

\begin{figure}[t]
    \centering
    \includegraphics[width=0.48\textwidth]{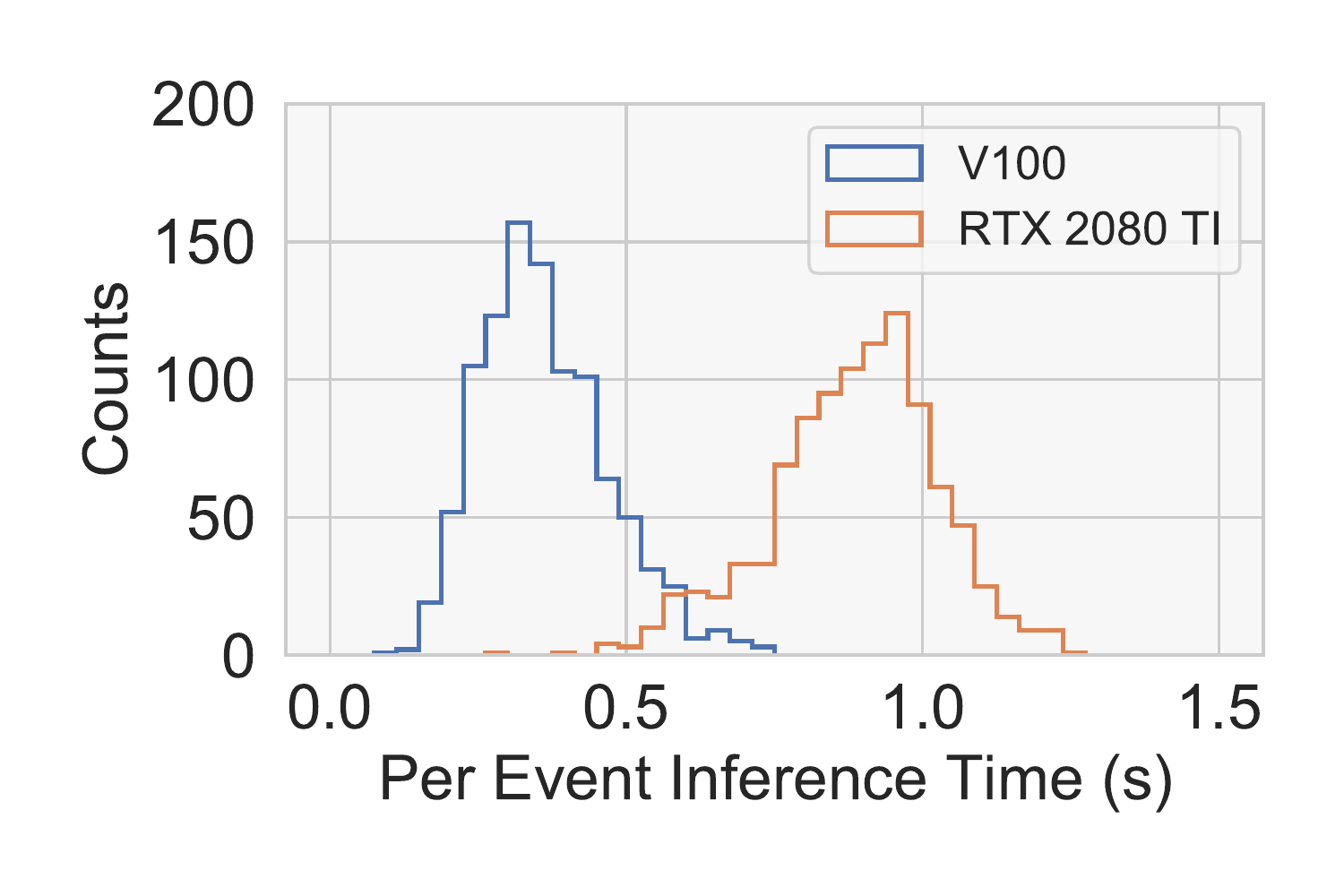}
    \caption{Comparison of inference wall-time per event in seconds. Post-processing time is not included.}
    \label{fig:time}
\end{figure}

All networks were trained on an NVIDIA V100 graphical processing unit (GPU), with approximately a week for training the full architecture. For the baseline model, we measured the per event inference time using the test set. 
All times are measured on a wall-clock. Figure~\ref{fig:time} shows the distribution of wall-clock time that it took for the network to compute the output using the best performing model. It compares two GPU models: the NVIDIA V100 and RTX 2080Ti, and does not include the time taken by the post-processing. Post-processing is performed on the CPU, and it takes 0.6~seconds on average per event.

\section{Conclusion}
We presented a region proposal-free deep learning algorithm applied to clustering of 3D pixels for reconstructing a particle trajectory. Using PILArNet, a public 3D LArTPC simulation dataset, we demonstrated that our algorithm achieves 98~\% in clustering efficiency and purity for track-like particles (HIP and MIP), and an efficiency of 93~\% and purity of 93~\% for EM showers. For all particles combined, our algorithm successfully clustered 90~\% of particles with an adjusted Rand index score greater than 92~\% with a mean pixel clustering efficiency and purity above 96~\%. 

The development of a high quality, full data reconstruction chain for LArTPC detectors remains a challenging research topic, and the accurate clustering of 3D pixels is a crucial step in the pipeline. Further, a machine learning based approach enables the full chain optimization by minimizing the error associated with all reconstruction target variables at the same time. Building on top of the previous study~\cite{laura}, we contribute to this development by addressing the challenging 3D pixel clustering step.

\vspace{0.6cm}
\section{Acknowledgement}
This work is supported by the U.S. Department of Energy, Office of Science, Office of High Energy Physics, and Early Career Research Program under Contract DE-AC02-76SF00515.

\bibliography{references}

\pagebreak

\makeatletter\onecolumngrid@push\makeatother
\newpage
\clearpage
\makeatletter\onecolumngrid@pop\makeatother

\onecolumngrid


\end{document}